\documentstyle[12pt]{article}
\begin{document}

\renewcommand{\baselinestretch}{1.5}
\newcommand\beq{\begin{equation}}
\newcommand\eeq{\end{equation}}
\newcommand\bea{\begin{eqnarray}}
\newcommand\eea{\end{eqnarray}}
\newcommand\fnx{\phi^x_n}
\newcommand\fnz{\phi^z_n}
\newcommand\fny{\phi^y_n}
\newcommand\cz{\chi^{zz}}
\newcommand\czq{\chi^{zz}(Q,0)}
\newcommand\czw{\cz(Q,\omega)}
\newcommand\imchi{\Im m \; \cz(Q,\omega)}
\newcommand\nbar{\overline{n}}

\centerline{\bf Majorana Fermion Representation For An Antiferromagnetic}
\centerline{\bf Spin-$1 \over 2$ Chain}
\vskip 1 true cm

\centerline{B. Sriram Shastry \footnote{{\it E-mail address}: 
bss@physics.iisc.ernet.in}}
\small
\centerline{\it Physics Department, Indian Institute of Science,}
\centerline{\it Bangalore 560012, India}
\centerline{also at}
\centerline{ \it Institute for Theoretical Physics,}
\centerline{ \it University of California, Santa Barbara 93116}
\vskip .75 true cm
\normalsize
\centerline{and}
\centerline{Diptiman Sen \footnote{{\it E-mail address}:
diptiman@cts.iisc.ernet.in}}
\small 
\centerline{\it Centre for Theoretical Studies, Indian Institute of Science,}
\centerline{\it Bangalore 560012, India}
\normalsize
\vskip 2 true cm

\noindent 
{\bf Abstract}
\vskip .5 true cm
We study the $1$-dimensional Heisenberg antiferromagnet with $s={1 \over 2}$ 
using a Majorana representation of the $s={1 \over 2}$ spins. A simple 
Hartree-Fock approximation of the resulting model gives a bilinear fermionic 
description of the model. This description is rotationally invariant and gives 
power-law correlations in the ``ground state'' in a natural fashion. The 
excitations are a two-parameter family of particles, which are spin-$1$ 
objects. These are contrasted to the ``spinon'' spectrum, and the technical 
aspects of the representation are discussed, including the problem of 
redundant states. 

\vskip 1 true cm
\noindent PACS numbers: ~75.10.Jm, ~75.50.Ee
 
\newpage

\centerline{\bf I. Introduction}
\vskip .5 true cm

The study of various representations of spins in terms of bosonic or fermionic
operators is an old and well studied problem, reviewed nicely, for example, in 
Ref. \cite{MATTIS}. The need for exploring various representations has 
received a further impetus from the recent interest in the Heisenberg
antiferromagnet, as a standard model in the Resonating Valence Bond theories 
\cite{PWA}, i.e., models where states with no long ranged N\'{e}el order
play an important role. The Schwinger boson representation is of very general 
validity, i.e., for any $s$, but the Schwinger fermionic representation is 
only valid for $s={1 \over 2}$ and gives $s^\alpha_i = \frac{1}{2} 
\sum_{\sigma,\sigma'} c^\dagger_{i \sigma} \tau^\alpha c_{i,\sigma'}$, with 
the constraint $\sum_\sigma c^\dagger_\sigma c_\sigma =1$ \cite{SCH,ABRIKOSOV}.
The constraint is not very easy to deal with, except in an averaged sense. 
Hence one may look for unconstrained representations. For $s={1 \over 2}$ such 
unconstrained representations can be found. The so called ``drone fermion'' 
representation \cite{KENAN,SPENCER,MATTIS} is one of the possibilities,
where we write $s_i^+= a^\dagger_i \phi_i$, $s_i^-= \phi_i a_i$ and
$s^z_i= a^\dagger_i a_i - \frac{1}{2}$, where the $a$'s are canonical
anticommuting variables, and $\phi_i$ is a {\em real} fermion with 
$\phi^\dagger=\phi$ and $\phi^2=1$. Thus $\phi$ is a `drone' whose only `job'
is to make spins at different sites commute, rather than anticommute. In 
single site problems like the Kondo problem, these are useful \cite{SPENCER}.
However, this representation violates rotation invariance, since
our choice of the z axis was arbitrary. A fully rotation invariant scheme
does exist, and can, for example, be derived from the above, by
simply rewriting the complex fermion $a$ in terms of its two
real components as $a\propto \phi_x+ i \phi_y$. This leads to a representation
with three Majorana fields, and is studied in this paper, in the context of
the $1$-dimensional Heisenberg model. 

The plan of this paper is as follows. In Sec. II, we discuss the Majorana
representation and the need for enlarging the Hilbert space of states in order
to obtain a representation of the Majorana algebra. We introduce the spin-$1
\over 2$ antiferromagnetic chain and its low-lying excitations in Sec. III.
In Sec. IV, we use the Majorana representation to study the chain within a 
rotationally invariant Hartree-Fock (H-F) approximation.
Since the H-F approximation is 
not unique in the general case, we require that the susceptibility calculated
by two methods, namely, from the energy change and from the 
fluctuation spectrum, should agree. This requirement, interestingly, rules out 
several possibilities, and leads to a particular scheme which is implemented.
We obtain a spectrum of low-lying excitations which bears a strong resemblance 
to the one discussed in Sec. III. 

We also discuss the spin of the Majorana fermion. In Sec. V, 
we compute the dynamic structure function
and susceptibility, at both zero and finite temperatures, and contrast 
these with previously known results. In Sec. VI, we study 
the response of the model to uniform and staggered magnetic fields. We end 
with some concluding remarks in Sec. VII.

\vskip .5 true cm
\centerline{\bf II. Majorana Representation}
\vskip .5 true cm

At each site $n$, we can write the spin operators ${\vec S}_n = {\vec 
\sigma}_n / 2$ in terms of three Majorana operators ${\vec \phi}_n$ as 
\cite{BSS,TSV,MAR}
\bea
\sigma^x_n ~&=&~ - ~i~ \phi^y_{n} ~\phi^z_{n} ~, \nonumber \\
\sigma^y_n ~&=&~ - ~i~ \phi^z_{n} ~\phi^x_{n} ~, \nonumber \\
{\rm and} \quad \sigma^z_{n} ~&=&~ - ~i~ \phi^x_{n} ~\phi^y_{n} ~.
\label{phi}
\eea
(We set Planck's constant equal to $1$). The operators ${ \phi}^a_{n}$ 
(with $a=x,y,z$) are hermitian and satisfy the anticommutation relations
\beq
\{ ~{ \phi}^a_{m} ~,~ { \phi}^b_{n}~ \} ~=~ 2~ \delta_{mn} ~ \delta_{ab} ~.
\label{antin}
\eeq
It is interesting to note that the relation ${\vec S}_n^2 = 3/4$ automatically
follows from Eqs. (\ref{phi}-\ref{antin}); one does not have to impose any
additional constraints at each site unlike the Schwinger representation 
\cite{SCH}. There is a local $Z_2$ gauge invariance since changing
the sign of ${\vec \phi}_n$ does not affect ${\vec S}_n$. (The Schwinger
representation has a local $U(1)$ gauge invariance).

For $N$ sites with a spin-$1 \over 2$ object at each site, the Hilbert space 
clearly has a dimension $2^N$. We now ask, what is the minimum possible 
dimension which will allow a representation of the form given in Eqs.
(\ref{phi}-\ref{antin})?  The answer is $2^{N+[N/2]}$, where $[N/2]$ denotes 
the largest integer less than or equal to $N/2$. This follows from the 
observation that a representation for (\ref{phi}-\ref{antin}) is given by 
\bea
\phi^a_{n} ~&=&~ \sigma^a_{n} ~\psi_n ~, \nonumber \\
{\rm where} \quad [ ~\sigma^a_{m} ~,~ \psi_n ~] ~&=&~ 0 ~, \nonumber \\
{\rm and} \quad \{ ~\psi_m ~,~ \psi_n ~\} ~&=&~ 2 ~\delta_{mn} ~.
\label{cliff}
\eea
The minimum dimension required for a matrix representation of the spinless
anticommuting operators $\psi_n$ is $2^{[N/2]}$ \cite{MAR}. Thus the Majorana 
representation of spin-$1 \over 2$ objects requires us to enlarge the space of 
states; the complete Hilbert space of states is given by a direct product of 
a `physical' space and an `unphysical' one. Now suppose that the Hamiltonian 
is purely a function of the physical operators
${\vec S}_n$; it therefore only acts on the physical states. Then the 
unphysical part of the Hilbert space simply factorizes out; hence each value 
of the energy will have a degeneracy of $2^{[N/2]}$. 

As an explicit example,
consider the case $N=2$. The Majorana Hilbert space is $8$-dimensional, where
the extra factor of $2$ arises from the unphysical space. We can denote the
$8$ states as $\uparrow \uparrow \uparrow$, $\uparrow \uparrow \downarrow$,
etc. The physical operators ${\vec S}_1$ and ${\vec S}_2$ only act on the
first and second symbols respectively. The third symbol, which may be 
$\uparrow$ or $\downarrow$, denotes the unphysical space. 
A Hamiltonian of the form ${\vec S}_1 \cdot {\vec S}_2$ only acts on the
first two symbols; hence the energy levels will be precisely the ones 
of a two-site antiferromagnet, but with an additional degeneracy of $2$
due to the third symbol. On the other hand, the Majorana 
operators can be written in the direct product form 
\bea
{\vec \phi}_1 ~&=&~ {\vec \sigma} \otimes 1 \otimes \sigma^x ~, \nonumber \\
{\rm and} \quad {\vec \phi}_2 ~&=&~ 1 \otimes {\vec \sigma} \otimes \sigma^y ~. 
\label{exam}
\eea
Hence they act on the third symbol and can therefore mix up physical and 
unphysical states.

One might worry that thermodynamic quantities like the entropy will get a 
spurious contribution proportional to $N$ due to the unphysical degeneracy 
of $2^{[N/2]}$. On the other hand,
when we make approximations like the H-F decomposition discussed 
later, the physical and unphysical states get mixed up in an essential way.
This completely changes the energy degeneracy; in particular, the H-F ground
state is actually unique as we will see.

We can think of $\phi_n^a$ as the fundamental field in our theory. Both
$\sigma_n^a$ and $\psi_n$ can be written in terms of $\phi_n^a$, as can be 
seen from Eq. (\ref{phi}) and $\psi_n = -i \phi_n^x \phi_n^y \phi_n^z$ 
respectively.

\vskip .5 true cm
\centerline{\bf III. Antiferromagnetic Spin-$1 \over 2$ Chain} 
\vskip .5 true cm

We will now begin our analysis of a Heisenberg antiferromagnetic chain. The 
Hamiltonian is
\beq
H ~=~ J ~\sum_n ~{\vec S}_n \cdot {\vec S}_{n+1} ~,
\label{ham}
\eeq
where the exchange constant $J > 0$. We use periodic boundary conditions
${\vec S}_{N+1} = {\vec S}_1$. (We set the lattice spacing $a=1$). 
The spectrum of (\ref{ham}) is exactly solvable by
the Bethe ansatz; in particular, the ground state energy is given by
$E_o = (- \ln 2 +1/4) N J$ $= -0.4431 NJ$. The lowest excitations are known to
be four-fold degenerate consisting of a triplet ($S=1$) and a singlet ($S=0$) 
\cite{FAD}. The excitation spectrum is described by a two-parameter continuum 
in the $(q,\omega)$ space, where $- \pi < q \le \pi$. The lower boundary of the 
continuum is described by the des Cloiseaux-Pearson relation \cite{dCP}
\beq
\omega_l (q) ~=~ {{\pi J} \over 2} ~\vert ~\sin ~q ~\vert ~,
\label{oml}
\eeq
whereas the upper boundary is given by
\beq
\omega_u (q) ~=~ \pi J ~\vert ~\sin ~{q \over 2} ~\vert ~.
\label{omu}
\eeq
We can understand this continuum by thinking of these excitations as being made
up of two spin-$1 \over 2$ objects ("spinons") with the dispersion \cite{FAD}
\beq
\omega (q) ~=~ {{\pi J} \over 2} ~\sin ~q ~,
\label{spinon}
\eeq
where $0 < q < \pi$.
A triplet (or a singlet) excitation with momentum $q$ is made up of two 
spinons with momenta $q_1$ and $q_2$, such that $0 < q_1 \le q_2 < \pi$, 
$q=q_1 + q_2$ if $0 < q \le \pi$, and $q = q_1 + q_2 - 2 \pi$ if 
$- \pi < q < 0$; further, $\omega (q) = \omega (q_1) +\omega (q_2)$. 
The two-parameter continuum arises because $q_1$ can vary from $0$ to $q/2$ 
if $0< q < \pi$, and from $\pi + q$ to $\pi + q/2$ if $- \pi < q < 0$. 

\vskip .5 true cm
\centerline{\bf IV. Hartree-Fock Treatment, Ground State And Excitations}
\vskip .5 true cm

We will now study this system using the Majorana representation. We write
(\ref{ham}) in terms of Majorana operators to get a quartic expression, and 
then perform a Hartree-Fock (H-F) decomposition. Thus we write:
\bea
H & = & ~-~  \frac{J}{4} ~\sum_n ~(~ \fnx \fny \phi^x_{n+1} \phi^y_{n+1} ~+~
\mbox{cycl. perm. (x,y,z)} ~) \nonumber \\
& \simeq & ~\frac{J}{4} ~\sum_n ~[~ \fnx \phi^x_{n+1} \langle \fny 
\phi^y_{n+1} \rangle ~+~
\langle  \fnx \phi^x_{n+1} \rangle \fny  \phi^y_{n+1} ~-~ \nonumber \\
& & \quad \quad \quad \quad \langle \fnx \phi^x_{n+1} \rangle \langle \fny 
\phi^y_{n+1} \rangle ~+~ \mbox{cycl. perm. (x,y,z)}~] ~.
\eea
In principle, the H-F can be done in three different ways; however rotational 
invariance implies that only one kind of bilinear can have a non-zero 
expectation value in the ground state. Namely,
\beq
 g = -i ~\langle ~\phi^a_{n} ~\phi^a_{n+1} ~\rangle ~,
\label{hf1}
\eeq
where $g$ has the same value for $a=x,y,z$; we also assume it to be 
translation invariant. The value of $g$ will be determined
self-consistently. We now have to diagonalize the quadratic Hamiltonian
\beq 
H ~=~ {{i J g} \over 2} ~\sum_{a,n} ~\phi^a_{n} ~\phi^a_{n+1} ~+~{3 \over 4}
N J g^2 ~.
\label{hfham1}
\eeq

Since $\phi^a_{n}$ is hermitian, its Fourier expansion can be defined as
\beq
\phi^a_{n} ~=~ {\sqrt{2 \over N}} ~\sum_{0<q<\pi} ~[~ b_{aq}^{\dag} ~
e^{iqn} ~+~ b_{aq}~ e^{-iqn} ~] ~,
\label{fourier}
\eeq
where 
\beq
\{ ~b_{aq} ~,~ b_{bq^{\prime}}^{\dag} ~\} ~=~\delta_{ab} ~\delta_{qq^\prime} ~. 
\label{antiq}
\eeq
A similar half zone definition of the fourier transforms is possible in higher 
dimensions as well; for example, on the square lattice, we could restrict the 
sum to $q_x >0$. We will work with {\it antiperiodic} boundary 
conditions for $\phi^a_{n}$ and {\it even} values of $N$ in order to eliminate 
modes with $q$ equal to $0$ and $\pi$. This simplifies the calculation because 
the momenta $q$ and $- q$ are then distinct points in the Brillouin zone 
extending from $- \pi$ to $\pi$. In Eq. (\ref{fourier}), $q=2 \pi (p - 1 /2) 
/N$, with $p= 1,2,...,N /2$. In the limit $N \rightarrow \infty$, we get 
\beq
H ~=~  ~\sum_{a} ~\sum_{0<q<\pi} ~\omega(q)~~b_{aq}^{\dag} b_{aq} ~+~ 
3 N J ~(~ {{g^2} \over 4}~ -~ {g \over \pi} ~)~,
\label{hfham2}
\eeq
where the Majorana fermions have the dispersion
\beq
\omega (q) ~=~ c ~\sin ~q ~,
\label{disp}
\eeq
 with $ c= 2 g J $. 
The H-F ground state $\vert ~0~ \rangle$ is therefore the state annihilated by 
all the $b_{aq}$. Note that it is unique unlike the {\it exact} ground state,
which has a degeneracy of $2^{N/2}$ within the Majorana formalism. It is 
curious that the H-F approximation gives a unique ground state which agrees
with the degeneracy we would have obtained {\it without} the Majorana 
formalism.

We now calculate (\ref{hf1}) in the H-F ground state and obtain
\beq
g ~=~ { 2 \over \pi} ~.
\label{g}
\eeq
The H-F ground state energy is therefore 
\beq 
E_{o~HF} ~=~ - ~{3 \over {\pi^2}} ~N J ~=~ - ~0.3040 ~N J ~.
\label{eo}
\eeq
This is greater than the exact value mentioned above; indeed, one can show
that {\it any} H-F decomposition must give an estimate for 
the ground state energy which is bounded below by the exact value $E_o$. The
argument goes as follows. In Sec. II, we have shown that the exact ground
state energy within the Majorana formalism is equal to the exact ground state
$E_o$ without the Majorana formalism, since the Hamiltonian $H$ only acts on 
physical states. Let us therefore prove the upper bound result in the Majorana 
Hilbert space which includes both physical and unphysical states. Now the H-F 
calculation is equivalent to self-consistently finding an ansatz ground state 
$\vert 0 \rangle$ and caclulating the expectation value of $H$ in that. (One 
can show that $\vert 0 \rangle$ is an eigenstate of the Majorana fermion number
operator. Hence an expectation value of the form $\langle ABCD \rangle$ is 
indeed given by the H-F decomposition $\langle AB \rangle \langle CD \rangle$ 
$- \langle AC \rangle \langle BD \rangle$ $+ \langle AD \rangle \langle BC 
\rangle$, if the operators $A$, $B$, $C$ and $D$ are all fermionic). By the 
variational argument, the expectation value of $H$ in any state is bounded 
below by $E_o$.

The "spinon" spectrum has the same form as in (\ref{spinon}) but has a 
different 
coefficient $c_{exact}= \pi J /2$, whereas we find $c= 4 J/ \pi$ from Eq. 
(\ref{g}). Note that the self consistent equation Eq. (\ref{hf1}) also 
leads to Eq. (\ref{g}), since we have
\beq
-i ~\sum_n ~\fnx \phi^x_{n+1} ~=~ \frac{2N}{\pi} ~-~ 4 ~\sum_{q > 0} ~
\sin q ~b^\dagger_{xq} b_{xq} ~. 
\label{useful}
\eeq

The ground state is a singlet since it is annihilated by the total spin
${\vec S}_{tot} = \sum_n {\vec S}_n$, for instance, by
\beq
S^z_{tot} ~=~ - i \sum_{0 < q < \pi} ~\Bigl( ~ b_{xq}^{\dag} b_{yq} ~-~ 
b_{yq}^{\dag} b_{xq} ~ \Bigr) ~.
\label{sz}
\eeq
We now ask: What is the spin of a Majorana fermion? From the commutation
relations between ${\vec S}$ and $b_{aq}^{\dag}$, we find that the one-fermion
state $b_{aq}^{\dag} ~\vert ~0~ \rangle$ has $S=1$. More specifically, the 
states $(b_{xq}^{\dag} + i b_{yq}^{\dag} ) ~\vert ~0~ \rangle$, 
$b_{zq}^{\dag} ~\vert ~ 0~ \rangle$, and $(b_{xq}^{\dag} - i b_{yq}^{\dag} ) ~
\vert ~0~ \rangle$ have $S^z =1, 0$ and $-1$ respectively.

A two-fermion state can therefore have $S=0,1$ or $2$ in general. However the 
state created by $S^z_{q} = 
\sum_n ~S^z_{n} e^{-iqn}~$, where $0 < q < \pi$, has the form 
\beq
S^z_{q} ~\vert ~0~ \rangle ~=~ -i ~\sum_{0 < k < q/2} ~\Bigl(~ b_{xk}^{\dag} 
b_{y,q-k}^{\dag} ~-~ b_{yk}^{\dag} b_{x,q-k}^{\dag} ~\Bigr) ~\vert ~0~ 
\rangle ~,
\label{sqz}
\eeq
and can be shown to have $S=1$. We have thus derived the two-parameter
continuum of triplet excitations in Eqs. (\ref{oml}-\ref{omu}), with a 
prefactor $4/ \pi$ instead of $\pi /2$.

Finally, we can compute the equal-time two-spin correlation function
\bea
G_n ~\equiv ~ \langle ~0~ \vert ~{\vec S}_o \cdot {\vec S}_n ~\vert ~0~ 
\rangle ~&=&~ {3 \over 4} \quad {\rm for} \quad n=0 ~, \nonumber \\
&=&~ -~ {3 \over {2 \pi^2 n^2}} ~[ 1 - (-1)^n ] \quad {\rm for} \quad n \ne 0~.
\label{corr}
\eea
This does not agree with the correct asymptotic behavior of $G_n$ which is
known to oscillate as $(-1)^n / n$. In particular, the H-F static structure 
function $S(q) = \sum_n G_n e^{-iqn}$ does not diverge as $q \rightarrow \pi$ 
in contrast to the correct $S(q)$ which has a logarithmic divergence at $\pi$.
Note that $\sum_n G_n =0$, as expected for a singlet ground state. It is 
interesting to observe that the Schwinger fermion representation yields a
correlation function which only differs from (\ref{corr}) by a numerical
factor (see the first reference in \cite{SCH}).

This Hartree-Fock state is readily generalized to finite temperatures, since 
we simply need to put in thermal population factors for the occupations of the 
fermions
\beq
\langle b^\dagger_{a q} b_{a q} \rangle = \frac{1}{1 + \exp (\beta c 
\sin q)} ~.
\label{thermal}
\eeq
Hence the self consistency condition Eq. (\ref{hf1}) together with
Eqs. (\ref{useful}) and (\ref{thermal}) gives us
\beq
g ~=~ \frac{2}{\pi} ~-~ \frac{4}{N} ~\sum_{q>0}~ 
\frac{\sin q}{1 + \exp (\beta c \sin q)} ~.
\eeq
It is easy to see that as $T \rightarrow \infty$ we have $g \rightarrow 0$,
and as $T \rightarrow 0$ we have $g  \rightarrow \frac{2}{\pi} 
( 1 -\frac{\pi^2 k_B^2 T^2}{6 c^2} )$, i.e., a power-law correction to the 
zero temperature `bandwidth' $g$.

The H-F ground state discussed above is, unfortunately, not the one 
with the lowest energy. If we allow a dimerized expectation value
$g_n$ in Eq. (\ref{hf1}), where $g_n$ can alternate in strength from bond
to bond, we find that the lowest energy is attained for the fully dimerized
state in which $g_n = 1$ for $n$ even and $0$ for $n$ odd (or vice versa).
This corresponds to a dimerized ground state with an energy 
\beq
E_{o~dim} ~=~ - {3 \over 8} N J ~,
\label{eodim}
\eeq
which is substantially lower than the earlier H-F value. There is a gap equal
to $J$ above the dimerized ground state. (This ground state is, of course, 
exact for the case $N=2$ \cite{COL}). The reader may wonder why we are
ignoring the dimerized H-F state in the rest of this paper, even though it
has the lowest H-F energy. The reason is that we know by other methods, both
analytical and numerical, that the correct ground state of the spin-$1 \over 2$
chain is translation invariant and that there is no gap above it. The H-F 
method is, after all, only an approximation, and different approximations 
can certainly give different results. We should therefore pick the 
H-F which agrees qualitatively with other methods; the ground state energy 
is not necessarily the best criterion for choosing one H-F over another. 
Having chosen 
a particular H-F on the basis of certain features, we of course have to check 
whether it reproduces other features equally well. We will see in Secs. V 
and VI that the translation invariant H-F yields reasonable results for 
the structure functions and susceptibilites also.

\vskip .5 true cm
\centerline{\bf V. Dynamic Structure Function and Susceptibility}
\vskip .5 true cm

We recall the definition of the dynamical susceptibility
\bea
\cz(Q,t) &=&  i \theta(t) ~\langle ~[~ S^z_{-Q} (t), S^z_Q ~]~ 
\rangle \\ \label{chiq}
\czw &=& \int_{-\infty}^{+\infty}\; dt ~\cz(Q,t) \; \exp(i \omega t) \\
&=& ~\sum_{\mu,\nu} ~\frac{ \exp(- \beta \epsilon_\nu) - 
\exp(- \beta \epsilon_\mu)} {\epsilon_\mu - \epsilon_\nu + \omega + i 0^+}~ 
<\mu | S^z_{-Q} | \nu > <\nu | S^z_Q | \mu> ~. \nonumber \\
\label{chi}
\eea
The Zeeman coupling of a spin to a magnetic field is given by $g_l \mu_B S^z 
B$, where $g_l$ and $\mu_B$ denote the Lande $g$-factor and the Bohr 
magneton respectively. The physical response
function (i.e. $g_l \mu_B <S^z>$) is $\chi = g_l^2 \mu_B^2 \czw$.
In the static limit $\omega=0$, we have the usual thermodynamic argument
for determining the susceptibility. If we perturb the system via
the coupling $H=H_0- g_l \mu_B B \sum_n \cos (Qn) S^z_n$,
then the change in the free energy is $\delta F = - g_l^2 \mu_B^2 B^2 \czq 
\theta_Q$, where $\theta_Q=1/4$ if $Q \neq 0 , \pi $, and $\theta_0 =1/2 =
\theta_\pi$. 
\footnote{This factor of $\theta$ arises because
for a finite $Q$ we drop two of the four terms in second order
perturbation theory using momentum conservation; this  neglect is
disallowed exactly at $Q=0, \; \pi$.}
Also recall that the static correlation function is given by
\beq
<S^z_{-Q} S^z_Q > ~=~ \int_{-\infty}^{+\infty} \;\; \frac{d\omega}{\pi}~
\frac{\imchi}{1- \exp(-\beta \omega)} ~. 
\label{statcorr}
\eeq
We will now compute the response functions in the H-F approximation.
We begin by expressing, for $0 < Q < \pi$, the operator $S_Q^z$ in terms 
of the Majorana fields in the Heisenberg picture:
\bea
S^z_Q (t) &=& -i \sum_{0<q<Q} \alpha(q,Q-q) ~b^\dagger_{xq} 
b^\dagger_{y,Q-q} ~\exp i (\omega_q + \omega_{Q-q}) t \nonumber \\
& & -i \sum_{\pi - Q<q< \pi} \alpha(q,2 \pi -Q-q) ~b_{xq} 
b_{y,2\pi -Q-q} ~\exp -i (\omega_q + \omega_{2 \pi -Q-q}) t \nonumber \\
& & -i \sum_{Q<q<\pi} \gamma(q,q-Q) ~[ b^\dagger_{xq} b_{y,q- Q} - 
b^\dagger_{yq} b_{x,q- Q}] ~\exp i (\omega_q - \omega_{q- Q})
t ~. \nonumber \\
\label{opeq}
\eea
In this equation we have introduced two real phenomenological functions
$\alpha(a,b)= \alpha(b,a) = \alpha (\pi -a, \pi -b)$ and $\gamma(a,b)$ 
which are, strictly speaking, equal to 
unity from the Majorana definition of the spins. These are introduced
in order to facilitate the comparison of our structure function with a 
phenomenological function proposed in Ref. \cite{MUL}. The essential point is
that we have assumed that the time evolution is given by the bilinear
in fermions, our Eq. (\ref{hfham2}). The representation for
$S^z_{-Q}$ is obtained by taking hermitean conjugates. Note that 
$S^z_Q$ or $S^z_{-Q}$ acting on the ground state generates two spinons. We 
insert it in Eq. (\ref{chiq}), carry out the contraction of the fermions
by Wick's theorem, and use Eq. (\ref{thermal}) in the form
$n_q=<b^\dagger_{q,\alpha} b_{q, \alpha} >$ and $\nbar_q= 1-n_q$ to find
\bea
\czw & = & \sum_{0<q<Q} \alpha^2(q,Q-q) ~\frac{\nbar_q \nbar_{Q-q} - 
n_q n_{Q-q}} {\omega_q + \omega_{Q-q} - \omega - i 0^+} \nonumber \\ 
& & ~+~ \sum_{0<q<Q} \alpha^2(q,Q-q) ~\frac{\nbar_q \nbar_{Q-q} - 
n_q n_{Q-q}} {\omega_q + \omega_{Q-q} + \omega + i 0^+}
\nonumber \\ 
& & ~+~ 2 ~\sum_{Q<q<\pi} \gamma^2(q,q-Q) ~\frac{n_{q-Q} \nbar_q- \nbar_{q-Q} 
n_q}{\omega_q - \omega_{q-Q} - \omega - i 0^+} ~. 
\label{chihf}
\eea
This is seen to be an even function of $\omega$ by using $q \rightarrow \pi + 
Q - q$ in the last term. Using Eq. (\ref{statcorr}), we deduce that
\bea
G^{zz}(Q) \equiv <S^z_{-Q} S^z_Q> &=& ~\sum_{0<q<Q} \alpha^2 (q,Q-q) [\nbar_q 
\nbar_{Q-q}  + n_q 
n_{Q-q}  ] \nonumber \\
& & +~ 2\sum_{Q<q<\pi} \gamma^2 (q,q-Q) n_{q-Q} \nbar_q ~.
\label{corrq}
\eea
Let us note that at zero temperature, if we set $\alpha= \gamma =1$, we get 
$G^{zz}(Q)= N |Q|/2 \pi$ and hence the correlation function quoted in Eq. 
(\ref{corr}). At the other extreme limit $T \rightarrow \infty$, we replace 
$n=\nbar=1/2$ and find $G^{zz}(Q)= N/4$. At {\it any} temperature, the 
relation $n_q + \nbar_q =1$ allows us to show that the sum rule 
$< S_n^z S_n^z > = 1/4$ is satisfied.

At zero temperature, we have the static susceptibility
\beq
\cz(Q,0) ~=~ 2 \sum_{0<q<Q} ~\frac{\alpha^2 (q,Q-q)}{\omega_q + \omega_{Q-q}} 
\eeq
which, in the standard situation $\alpha=1$, can be evaluated in the closed 
form
\beq
\cz(Q,0) ~=~ \frac{N}{\pi c \sin (Q/2)} ~\log ~[~ \frac{\cos ~(\pi-Q)/4}
{\cos ~(\pi + Q)/4} ~]~.
\eeq
The uniform value is 
\beq 
\cz(0,0) ~=~ \frac{N}{\pi c} ~=~ \frac{N}{4J}
\label{chiuniform}.
\eeq
The neutron scattering function which is of particular interest is found at 
zero temperature as
\beq
\imchi ~=~ \pi ~\sum_{0<q<Q} ~\alpha^2 (q,Q-q) ~\delta \ (\omega_q + 
\omega_{Q-q} - \omega) 
\label{neutron}.
\eeq
for $\omega > 0$. We can evaluate it in terms of the dimensionless energies
$u \equiv \omega/c$, $u_> \equiv 2 \sin (Q/2)$ and $u_< \equiv \sin Q$, as 
\beq
\imchi = \frac{N}{c} \frac{\alpha^2(q^*,Q-q^*)}{|\cos(q^*) - \cos(Q-q^*)|}
\theta(u_>-u) \; \theta(u - u_<) \label{phen1}
\eeq
where $q^*$ is the solution of $\sin  q^* + \sin (Q-q^*)=u$ which equals
$Q/2$ at $u= u_>$. With this we find
\bea
\sin q^* ~&=&~ \frac{1}{2} ~[~ u - \cot (Q/2) \sqrt{u_>^2-u^2} ~] \nonumber \\
\cos q^* ~&=&~ \frac{1}{2} ~[~ u \cot (Q/2) + \sqrt{u_>^2-u^2} ~]~. 
\label{phen2}
\eea
This implies that $|\cos(q^*)-\cos(Q-q^*)|=\sqrt{u^2_>-u^2}$, and 
\beq
\imchi ~=~ \frac{N}{c} ~\frac{\alpha^2(q^*,Q-q^*)}{\sqrt{u^2_>-u^2}} ~
\theta(u_>-u) \; \theta(u - u_<) ~.
\label{phen3}
\eeq
This susceptibility is very similar to that proposed in Ref. \cite{MUL} 
phenomenologically, and also found for the long ranged spin-$1 \over 2$ chain 
\cite{HAL,SHAS} in Ref. \cite{HZIRN}, with one important difference. The 
spectral weight here is dominatd by the upper threshold of the two parameter 
continuum $u_>$, whereas the weight is peaked at the lower threshold $u_<$ in 
Ref. \cite{MUL}. It is straightforward to see that if we choose
\beq
\alpha^2 (q,Q-q) ~\equiv ~\nu ~\frac{|\sin (Q/2 -q)|}{\sqrt{\sin q} \;
\sqrt{\sin (Q-q)} } ~,
\label{phen4}
\eeq
then on using Eq. (\ref{phen2}), the weight is shifted to the bottom, and we 
get
\beq
\imchi ~=~ \frac{N \nu }{c} ~\frac{1}{ \sqrt{u^2-u_{<}^2} } ~\theta(u_>-u) 
\; \theta(u - u_<) ~.
\label{phen5}
\eeq
With this choice, the static correlation function can be evaluated
from Eq. (\ref{statcorr}). We find 
\beq
G^{zz}(Q) ~=~ \frac{N \nu}{\pi} ~\log ~[~ \frac{1+ \sin (Q/2)}{\cos (Q/2)} ~]~,
\label{phen6}
\eeq
leading to the asymptotic behaviour $\sim (-1)^n/n$ at long distances.
Indeed one can use the two parameters $c$ and $\nu$ in Eqs. (\ref{phen5}-
\ref{phen6}) together with the various sum rules known, in order to obtain 
very realistic structure functions which mimic the behaviour of the nearest 
neighbour Heisenberg antiferromagnet. At finite temperatures, we find from Eq. 
(\ref{corrq}) in the usual case of $\alpha =\gamma =1$
\beq
<S^z_n S^z_0>= \frac{1}{4} \delta_{r,0} - {1 \over {16}} ~[f_n ({{\beta c} 
\over 2})]^2 ~,
\eeq
with
\beq
f_n ({{\beta c} \over 2}) ~=~ {2 \over {\pi}} ~\int_o^{\pi} ~dx ~\sin (nx) ~
\tanh ({{\beta c} \over 2} \sin x) ~,
\eeq
leading to an exponentially decaying correlation function with a correlation 
length $\xi \sim 1/T$ for $T \rightarrow 0$. The function $f_n$ vanishes
for even $n$ in contrast to one's usual expectation.  In the presence of the
phenomenological $\alpha$, one must necessarily cut off the linear
divergence of $\alpha$ at $Q=\pi$ and $q \sim 0,\pi$. A temperature dependent 
cutoff, such as $\alpha^2(a,b)= ( |\sin(a-b)/2| + (\mbox{const})^2 T)/(~\sqrt {
\sin(a) + (\mbox{const}) T} \sqrt{ \sin(b) + (\mbox{const}) T}~)$ interpolates 
nicely between the zero temperature limit and the high temperature limit, and 
again gives a correlation length $\sim 1/T$. 

\vskip .5 true cm
\centerline{\bf VI. Magnetic Fields}
\vskip .5 true cm

We will now discuss the H-F ground state of the spin chain in the presence of
uniform and staggered magnetic fields, and calculate the two susceptibilities.

\vskip .5 true cm
\centerline{\bf A. Uniform Magnetic Field}
\vskip .5 true cm

For an uniform magnetic field $B {\hat z}$, we add a term $-g_l \mu_B B \sum_n 
S^z_{n}$ to the Hamiltonian (\ref{ham}). Since this term commutes with 
(\ref{ham}), we can use the same H-F decomposition as in (\ref{hf1}) with 
$g = 2 / \pi$. Since the extra term in the Hamiltonian is quadratic in the 
Majorana operators, we only have to perform a rediagonalization of 
(\ref{hfham1}). We find that modes with $S^z = \pm 1$ have an energy
\beq
\omega_{\pm} (q) ~=~ {{4 J} \over \pi} \sin ~q ~\mp~ g_l \mu_B B ~,
\label{ompm}
\eeq
while the energy of the $S^z = 0$ modes remain unchanged. For $B > 0$, let us 
define a momentum $q_o$ such that 
\beq
q_o ~=~ \sin^{-1} ~( {{\pi g_l \mu_B B} \over {4J}} ) ~,
\label{qo1}
\eeq
and $0 < q_o < \pi /2$. (Such a $q_o$ exists only if the magnetic field is 
less than a critical value $B_c = 4 J / \pi g_l \mu_B $). Then the modes with 
$S^z =1$ and momenta lying in the range $0 < q < q_o$ and $\pi - q_o < q < 
\pi$ have negative energy, and the ground state of the system is one in which 
those modes are occupied. The change in the ground state energy is therefore 
given by a sum over all the occupied modes $q$,
\bea
\Delta E_{o~HF} ~&=&~ \sum_q ~\Bigl( ~{{4J} \over \pi} ~\sin ~q ~-~ g_l \mu_B 
B ~ \Bigr) \nonumber \\
&=&~ {{4NJ} \over {\pi^2}} ~( 1 - \cos q_o) ~-~ {{N g_l \mu_B B} \over \pi} ~
q_o ~.
\label{eounif}
\eea
The expectation value of $S^z$ in the ground state is obtained either by 
counting the number of occupied modes, or by differentiating (\ref{eounif})
with respect to $g_l \mu_B B$. Thus
\beq
\langle ~S^z ~\rangle~=~ {{N q_o} \over \pi} ~=~ {N \over \pi} ~\sin^{-1} 
\Bigl(~ {{\pi g_l \mu_B B} \over {4J}} ~ \Bigr) ~.
\label{szunif}
\eeq
Finally, the (uniform) susceptibility is given by
\beq
\chi ~=~ {1 \over {g_l \mu_B}} ~\Bigl( {{\partial ~\langle S^z \rangle} \over 
{\partial B}} ~ \Bigr)_{B=0} ~=~ {N \over {4J}} ~.
\label{unifsus}
\eeq
This agrees with the result in the previous section. For a strong magnetic
field $B > B_c$, the ground state is fully polarized with $S^z = N/2$. These
results are to be compared with the
exact results for the susceptibility $\chi = N / \pi^2 J$, and the 
critical field $B_c = 2 J / g_l \mu_B$ \cite{MUL}.

Since $S^z_{n}$ has a non-zero expectation value in the ground state, 
the above calculation is not entirely self-consistent, i.e., one
should also allow H-F decompositions of the form 
\bea
\langle ~\phi^x_{n} ~\phi^y_{n} ~\rangle ~&=&~ i f_o ~, \nonumber \\ 
{\rm and} \quad \langle ~\phi^x_{n} ~\phi^y_{n \pm 1} ~\rangle ~&=&~ i 
f_{\pm 1} ~. 
\label{hf2}
\eea
Further, the expectation values 
\bea
\langle ~\phi^x_{n} ~\phi^x_{n+1} ~\rangle ~=~ \langle ~\phi^y_{n} ~
\phi^y_{n+1} ~\rangle ~&=&~ i g_T ~, \nonumber \\
{\rm and} \quad
\langle ~\phi^z_{n} ~\phi^z_{n+1} ~\rangle ~&=&~ i g_L
\label{hf3}
\eea
may be unequal since the magnetic
field breaks rotational invariance. On doing this more general H-F calculation, 
we find that although the ground state remains the same qualitatively (i.e.,
a number of $S^z =1$ modes have to be filled in the regions $0 < q < q_o$ and
$\pi - q_o < q < \pi$), various numbers change. For instance, $q_o$ is now 
given by
\beq
q_o ~+~ \sin ~q_o ~(1 + \cos ~q_o ) ~=~ {{\pi g_l \mu_B B} \over {2J}} ~.
\label{qo2}
\eeq
The H-F parameters are 
\bea
g_T ~&=&~ {2 \over \pi} ~\cos ~q_o ~, \quad g_L ~=~ {2 \over \pi} ~, 
\nonumber \\
f_o ~&=&~ {{2 q_o} \over \pi} ~, \quad f_{\pm 1} ~=~ 0 ~.
\label{hf4}
\eea
Since the magnetization is equal to $N q_o / \pi$, the susceptibilty is $\chi 
= N /6J$. (The critical field for complete polarization is $B_c = J ( 1 + 2/ 
\pi )/ g_l \mu_B$). We therefore have the curious result that a completely
self-consistent H-F calculation does not agree with linear response theory for 
small fields.

\vskip .5 true cm
\centerline{\bf B. Staggered Magnetic Field}
\vskip .5 true cm

We now study the situation with a staggered magnetic field. We add a term
$- g_l \mu_B B \sum_n (-1)^n S^z_{n}$ to the Hamiltonian and perform a H-F 
decomposition. As in the uniform case, we will assume that $g_T = g_L = 2 / 
\pi$ and $f_o = f_{\pm 1} = 0$ in Eqs. (\ref{hf2}-\ref{hf3}) even though this 
is not completely self-consistent. We then find that the dispersion of the 
longitudinal modes remain the same as before while those of the transverse 
modes change. To be explicit,
\bea
\omega_L (q) ~&=&~ {{4J} \over \pi} ~ \sin ~q ~, \nonumber \\
{\rm and} \quad \omega_T (q) ~&=&~ \Bigl( ~{{16 J^2} \over {\pi^2}} ~\sin^2 ~
q ~+~ g_l^2 \mu_B^2 B^2 ~ \Bigr)^{1/2} ~.
\label{omlt}
\eea
Further, the change in the ground state energy is
\beq
\Delta E_{o~HF} ~=~ \sum_{0 < q < \pi} ~\Bigl( ~{{4J} \over \pi} ~\sin ~q ~-~ 
\omega_T (q) ~\Bigr) ~.
\label{eostag}
\eeq
On differentiating this with respect to $g_l \mu_B B$, we find the staggered 
magnetization to be 
\beq
\langle ~\sum_n ~(-1)^n ~S^z_{n} ~\rangle ~=~ N g_l \mu_B B ~\int_0^{2 \pi} ~
{{dq} \over {2 \pi}} ~{1 \over \omega_T (q)} ~.
\label{szstag}
\eeq
For small fields, this goes as $(N g_l \mu_B
B /4J) \ln (J / g_l \mu_B B)$ which implies that the staggered susceptibility 
is divergent. This is the correct result. For large fields, the staggered
magnetization approaches $N/2$ as it should.

\vskip .5 true cm
\centerline{\bf VII. Discussion}
\vskip .5 true cm

To summarize, we have used a Majorana fermion representation to study a 
nearest-neighbor isotropic antiferromagnetic spin-$1 \over 2$ chain. Within
a translation invariant Hartree-Fock approximation, we have found the spectrum
of low-lying excitations, the two-spin correlation function, the structure
function, and the magnetic susceptibilities. All of these agree qualitatively
with the results found earlier by a variety of other methods. The agreement
can be made quantitative if we introduce some phenomenological functions within
the Majorana formalism.

It is somewhat surprising that a fully dimerized Hartree-Fock approximation 
leads to a ground state with a lower energy. One way of stabilizing the 
translation invariant ground state with respect to the dimerized one is to 
apply an uniform magnetic field with a strength $B > 0.5829 B_c = 0.7422 J / 
g_l \mu_B $. Such a magnetic field lowers the energy of the translation 
invariant ground state below $-3NJ/8$, and does not change the energy of the 
dimerized ground state, for $B < J/g_l \mu_B$, due to the finite gap to spin 
excitations.

It would be interesting to go beyond our Hartree-Fock treatment and study the
effects of fluctuations. Besides producing more accurate numbers for various
quantities such as the spin wave velocity, such a study could also lead to a 
more detailed understanding of the "spinons" in a spin-$1 \over 2$ chain 
in terms of Majorana fermions.

It may be instructive to examine models with anisotropy, frustration, and
higher dimensionality using the Majorana representation, and to compare 
with known results. Amongst other things, this would help to determine the 
range of validity of this way of studying spin-$1 \over 2$ systems. 

We have briefly examined the ferromagnetic case in which the exchange constant
in Eq. (\ref{ham}) is {\it negative}. We perform a non-rotation invariant 
Hartree-Fock decomposition by allowing $\sigma_n^z = -i \phi_n^x \phi_n^y$ to 
take an expectation value. We then obtain the correct ground state energy
$E_o = NJ /4$, with the total $S^z = \pm N/2$. However we get the wrong
dispersion relation, including a gap, for the low-energy excitations. Thus the
Majorana Hartree-Fock approximation is not a good starting point for studying 
the spin-$1 \over 2$ ferromagnet.

\vskip .5 true cm
\centerline{\bf Acknowldgements}
\vskip .5 true cm

B. S. S. would like to thank several colleagues who have encouraged him to 
publish Ref. \cite{BSS}, particularly P. W. Anderson, G. Baskaran and R. 
Shankar, which  represented  a preliminary version of this work.
This work was supported in part by the National Science Foundation under Grant 
No. PHY94-07194.

\end{document}